# Generalized Rules of Coherence Transfer from Local to Global Scale

Moses Fayngold
*Department of Physics, New Jersey Institute of Technology, Newark, NJ 07102*

A thought experiment with the path-entangled photon pairs is suggested. Its analysis predicts elimination of local coherence even at infinitesimally weak entanglement. Local coherence turns out to be totally incompatible with entanglement. We can thus predict and name a new phenomenon – "total mutual intolerance" between local and global coherence. Unlike incompatible observables like position and momentum, whose expectation values can still coexist under trade-off between the respective indeterminacies, there is no coexistence between local and global coherence. This prediction, if confirmed, may open some new venues in Quantum Physics and Quantum Information theory.

Key words: Bi-photon, entanglement, correlations, coherence transfer

## Introduction

This work describes a thought experiment with a path-entangled photon pair, which allows one to vary the entanglement strength and characteristics of local superposition.

The first section reviews briefly the experiments by J. G Rarity, P. R. Tapster [1] and Z. Ou, X. Zou, L. Wang, L. Mandel [2]. Following A. Hobson [3-5], we will refer to their experiments as RTO. In the RTO, the pair of maximally entangled photons A and B (bi-photon AB) was produced by a common source S. The two pathways labeled as 1 and 2 formed two momentum eigenstates $|1\rangle$ and $|2\rangle$. When disentangled, each photon can be in their superposition and accordingly interfere with itself. But entanglement eliminates their local interference.

Sec.2 describes the suggested generalized scheme which allows one to use asymmetric beam splitters (BS) and most important, to monitor the entanglement strength from maximal to zero. The analysis unveils an unexpected new feature in the local-global scale correlations: the local coherence turns out to be totally incompatible with system's entanglement, no matter how weak. This can be named the "Mutual intolerance" effect which may also be observable for entangled electron pairs [6].

The basic results are summarized in Conclusion.

## 1. The RTO experiments

The RTO experiments showed *how entanglement between two photons A and B affects the photon self-interference. Each photon* in the pair (AB) can move along either of the two paths - path 1 (the solid line) and path 2 (the dashed line). Mirrors M reflect the paths directly to the respective BS. Sides A and B work as back to back Mach-Zehnder interferometers (Fig.1).

(In the actual RTO arrangement the photons were emitted into two angular cones, but the simplified representation used in [3-5] and here does not affect the results).

When independent, each photon A and B can be in a superposition

$$|\Psi\rangle_A = \frac{1}{\sqrt{2}}\left(|1\rangle_A + e^{i\phi_A}|2\rangle_A\right), \quad |\Psi\rangle_B = \frac{1}{\sqrt{2}}\left(|1\rangle_B + e^{i\phi_B}|2\rangle_B\right) \tag{1.1}$$

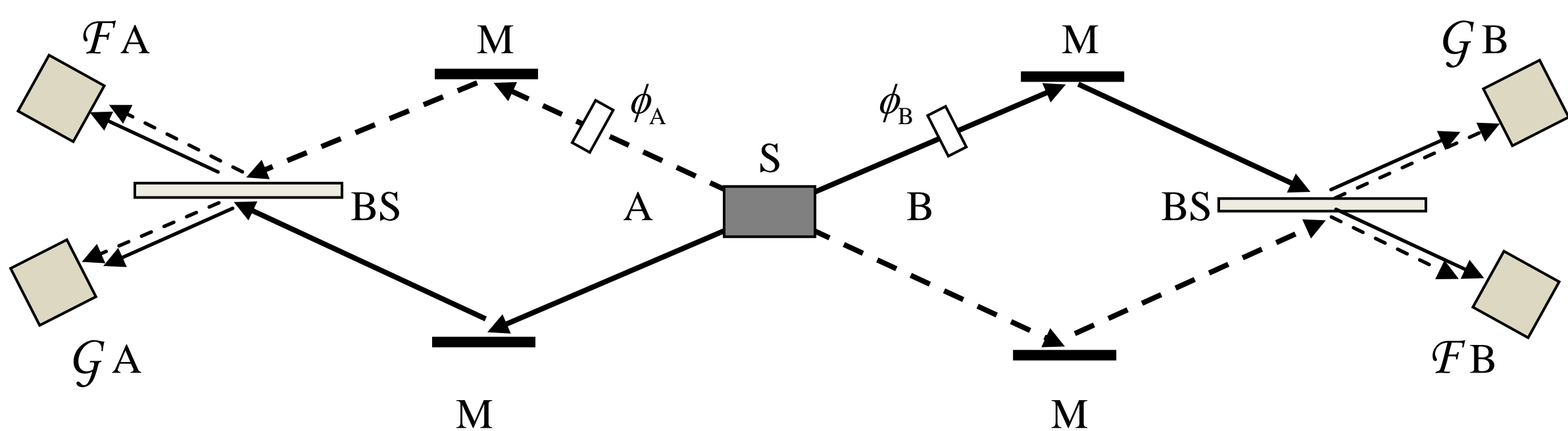

**Fig. 1** Setup of the RTO experiments with a bi-photon AB from the source S

The BS-s recombine the beams, so each photon can interfere with itself at the respective station. The phase-shifters $\phi_A$ and $\phi_B$ monitor the interference pattern. Detectors $\mathcal{F}$A, $\mathcal{G}$A, $\mathcal{F}$B, $\mathcal{G}$B record the corresponding arrivals.

Focusing on RTO, we write for *entangled* bi-photon after passing through phase-shifters:

$$|\Psi\rangle_{AB} = \frac{1}{\sqrt{2}}\left(e^{i\phi_B}|1,1\rangle + e^{i\phi_A}|2,2\rangle\right) = \frac{e^{i\phi_B}}{\sqrt{2}}\left(|1,1\rangle + e^{i\phi}|2,2\rangle\right), \quad \phi \equiv \phi_A - \phi_B \tag{1.2}$$

Here each double ket is the shorthand for the respective direct product $|j\rangle_A |j\rangle_B \equiv |j,j\rangle$, $j = 1, 2$, with the first argument standing for A and the second one – for B.

The photon state $|\mathcal{F}\rangle_A$ activating $\mathcal{F}$A is a superposition of transmitted part of $|1\rangle_A$ and reflected part of $|2\rangle_A$. Similarly, state $|\mathcal{G}\rangle_A$ activating $\mathcal{G}$A is the superposition of reflected part of $|1\rangle_A$ and transmitted part of $|2\rangle_A$. Each such state is represented in Fig.1 by a pair of parallel arrows – solid and dashed. And the same holds for the B-photon. Measurements of state (1.2) in the $\left(|\mathcal{G}\rangle, |\mathcal{F}\rangle\right)$-basis may show, apart from (+) correlations (activating the equally-labeled detectors), also $(-)$ correlations (activating differently-labeled detectors). The corresponding probabilities in the RTO case (symmetric BS and maximally strong entanglement) are given by:

$$\mathcal{P}^+(\mathcal{F},\mathcal{F}) = \mathcal{P}^+(\mathcal{G},\mathcal{G}) = \frac{1}{4}\left(1 + \cos(\alpha + \phi)\right), \tag{1.3}$$

$$\mathcal{P}^-(\mathcal{F},\mathcal{G}) = \mathcal{P}^-(\mathcal{G},\mathcal{F}) = \frac{1}{4}\left(1 - \cos(\alpha + \phi)\right), \tag{1.4}$$

where $\alpha$ includes additional phases due to elements other than phase-shifters. The periodic terms here show the *nonlocal* interference [1, 2, 7, 8]. But the sum of (1.3) and (1.4) giving the *local* probability $\mathcal{P}(\mathcal{F})$ is phase independent, showing elimination of the local coherence.

## 2. Generalized scheme: bi-photon in an arbitrary basis

Here we describe the thought experiment opening a far broader view of the whole phenomenon. First, we insert an absorbing plate (AP) in one of the paths, say, path 1 on the A side (Fig.2). This will extend case (1.1) for photon A to

$$|\Psi\rangle_A = \frac{1}{\sqrt{2}}\left( \tilde{p}\,|1\rangle_A + e^{i\phi_A}|2\rangle_A \right) \qquad (2.1)$$

Hereafter, a symbol with tilde will denote complex variable, so $\tilde{p} = pe^{i\xi}$, and $p$ can be considered real positive. Physically, $\tilde{p}$ is the transmission amplitude of AP, $0 \le p^2 \le 1$.

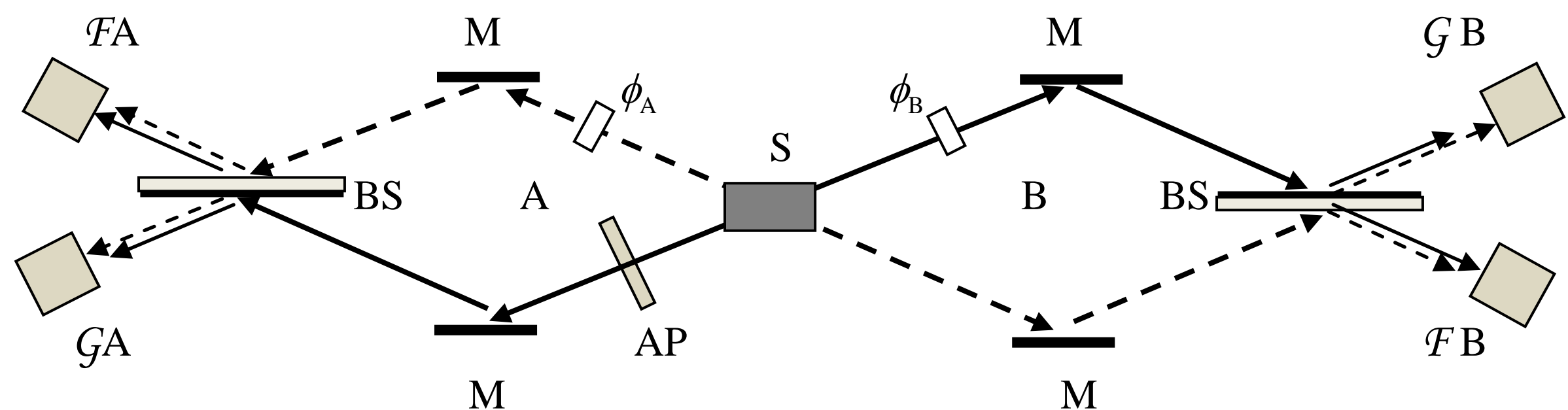

**Fig. 2** The setup for a thought experiment generalizing RTO.
The inserted plate AP monitors the ratio of superposition amplitudes for paths 1 and 2. Farther monitoring is achieved by allowing BS to be asymmetric.

The plate should be non-reflective to exclude the chance of sending A to the B side. Using only one plate makes (2.1) and the following equations asymmetric, but the corresponding Math is simpler without changing the results.

The inserted AP reduces probability to find A on the post-AP stretch to $\mathcal{P}_1 = p^2/2$ vs. $\mathcal{P}_2 = 1/2$ for path 2, so the relative beam strengths are described by the ratio

$$\mathcal{P}_1 / \mathcal{P}_2 \equiv \varepsilon = p^2 \qquad (2.2)$$

There will accordingly emerge a new possible outcome (photon absorption, no detector clicks on the A side). Denote the resulting vacuum state as $|0\rangle$, and the respective probability as $\mathcal{P}_0$. Then

$$\mathcal{P}_0 = (1 - p^2)/2 \qquad (2.3)$$

The Hilbert space $\mathcal{H}$ of the system expands to three dimensions, and $\mathcal{P}_0+\mathcal{P}_1+\mathcal{P}_2=1$. But we will use only a 2D basis $(|1\rangle,|2\rangle)$ or its rotated version $(|\mathcal{F}\rangle,|\mathcal{G}\rangle)$. All this would hold for photon B as well if we choose to insert the AP on the B-side or if the bi-photon is entangled.

For the latter case, consulting with Fig.2, we generalize (1.2) to

$$|\Psi\rangle_{AB}=\frac{1}{\sqrt{2}}\left\{\tilde{p}\,|1,1\rangle+e^{i(\alpha+\phi)}\,|2,2\rangle\right\} \tag{2.4}$$

The amplitude $\tilde{p}=pe^{i\xi}$ in (2.3) is the same as in (2.1) *if* AB is path-entangled and only the events with detector clicks on both sides are counted. Physically, (2.3) describes the system within time interval corresponding to photon traveling *between* AP and BS.

The second generalization involves BS. Assuming them identical like in [3-5], we allow each to be asymmetric [9, 10] (see also [11-15]). Let orient them such that the reflective side of each BS faces path 1. Denote their transmission and reflection amplitudes as $\tilde{t}\equiv te^{i\tau}$, $\tilde{r}\equiv re^{i\rho}$, respectively. Then we can write transformation $(|1\rangle,|2\rangle)\Rightarrow(|\mathcal{M}\rangle,|\mathcal{N}\rangle)$ for each photon in two steps. First,

$$\begin{pmatrix}|\mathcal{F}\rangle\\|\mathcal{G}\rangle\end{pmatrix}=\mathcal{R}\begin{pmatrix}|1\rangle\\|2\rangle\end{pmatrix},\quad\text{with}\quad\mathcal{R}=\begin{pmatrix}\tilde{t}&\tilde{r}\\\tilde{r}&\tilde{t}\end{pmatrix} \tag{2.5}$$

Unitarity imposes the restraints

$$t^2+r^2=1;\qquad\tau-\rho=\pi/2, \tag{2.6}$$

so the determinant of $\mathcal{R}$ is $D(\mathcal{R})=\tilde{t}^2-\tilde{r}^2=e^{2i\tau}=-e^{2i\rho}$.

Then, discarding the immaterial factor $D^{-1}(\mathcal{R})$ we can write the inverse transformation as

$$\begin{pmatrix}|1\rangle\\|2\rangle\end{pmatrix}=\mathcal{R}^{-1}\begin{pmatrix}|\mathcal{F}\rangle\\|\mathcal{G}\rangle\end{pmatrix}=\begin{pmatrix}\tilde{t}&-\tilde{r}\\-\tilde{r}&\tilde{t}\end{pmatrix}\begin{pmatrix}|\mathcal{F}\rangle\\|\mathcal{G}\rangle\end{pmatrix} \tag{2.7}$$

Consulting with Fig.2 and putting (2.7) into (2.4) yields

$$|\Psi\rangle_{AB}=\frac{1}{\sqrt{2}}\left\{\left(\tilde{p}\tilde{t}^2+\tilde{r}^2\right)|\mathcal{F},\mathcal{F}\rangle+\left(\tilde{p}\tilde{r}^2+\tilde{t}^2\right)|\mathcal{G},\mathcal{G}\rangle-\left(\tilde{p}+1\right)\tilde{r}\tilde{t}\left(|\mathcal{F},\mathcal{G}\rangle+|\mathcal{G},\mathcal{F}\rangle\right)\right\} \tag{2.8}$$

The general expression (2.8) shows a "change of face" of the entanglement [6]**:** the $(+)$ correlated state (2.4) is a superposition of $(+)$ and $(-)$ correlations in the $(|\mathcal{F}\rangle,|\mathcal{G}\rangle)$ basis.

We can reduce the number of variables here by introducing the ratio $\eta\equiv t^2/r^2$, so that

$$t^2 = \eta/(1+\eta)\,; \quad r^2 = 1/(1+\eta) \tag{2.9}$$

According to (2.2, 6, 9), we have for probabilities of the two $(+)$ correlations

$$\mathcal{P}(\mathcal{F}, \mathcal{F}) = \frac{1}{2}\left|\left(\tilde{p}\tilde{t}^2 + \tilde{r}^2\right)\right|^2 = \frac{1+\varepsilon\eta^2 + 2\sqrt{\varepsilon}\,\eta\cos w}{2(1+\eta)^2} \tag{2.10}$$

and

$$\mathcal{P}(\mathcal{G},\mathcal{G}) = \frac{1}{2}\left|\left(\tilde{p}\tilde{r}^2 + \tilde{t}^2\right)\right|^2 = \frac{\varepsilon+\eta^2 + 2\sqrt{\varepsilon}\,\eta\cos w}{2(1+\eta)^2}\,, \quad w \equiv \phi + \xi \tag{2.11}$$

As seen from (2.10, 11), $\mathcal{P}(\mathcal{F},\mathcal{F}) \neq \mathcal{P}(\mathcal{G},\mathcal{G})$. The total probability of $(+)$ correlations is

$$\mathcal{P}^+ \equiv \mathcal{P}(\mathcal{F}, \mathcal{F}) + \mathcal{P}(\mathcal{G}, \mathcal{G}) = \frac{(1+\varepsilon)(1+\eta^2) + 4\eta\sqrt{\varepsilon}\cos w}{2(1+\eta)^2} \tag{2.12}$$

The probabilities for $(-)$ correlations also obtain from (2.8) and are equal to each other, with their sum

$$\mathcal{P}^- \equiv \mathcal{P}(\mathcal{F}, \mathcal{G}) + \mathcal{P}(\mathcal{G}, \mathcal{F}) = \frac{\eta}{(1+\eta)^2}\left(1+\varepsilon - 2\sqrt{\varepsilon}\cos w\right) \tag{2.13}$$

Periodic terms in (2.10-13) are the hallmark of two-particle interference. The expressions (2.12, 13) are equivalent to (1.3, 4) at $\varepsilon = \eta = 1$. But their sum $\mathcal{P}^+ + \mathcal{P}^-$ is less than 1 in the above-mentioned 2D subspace of $\mathcal{H}$. Adding $\mathcal{P}_0$ from (2.3) will give $\mathcal{P}_0 + \mathcal{P}^+ + \mathcal{P}^- = 1$.

The visibility of patterns (2.12, 13) defined as the ratio of the amplitude of periodic term to the constant term will be

$$\mathcal{V}^+(\varepsilon, \eta) = \frac{4\sqrt{\varepsilon}\eta}{(1+\varepsilon)(1+\eta^2)} \quad \text{for } \mathcal{P}^+ \tag{2.14}$$

and

$$\mathcal{V}^-(\varepsilon) = \frac{2\sqrt{\varepsilon}}{1+\varepsilon} \qquad \text{for } \mathcal{P}^- \tag{2.15}$$

The $\mathcal{V}^-(\varepsilon)$ is independent of characteristics of BS. Both cases describe nonlocal interference of a bi-photon and are shown graphically in Fig.3.

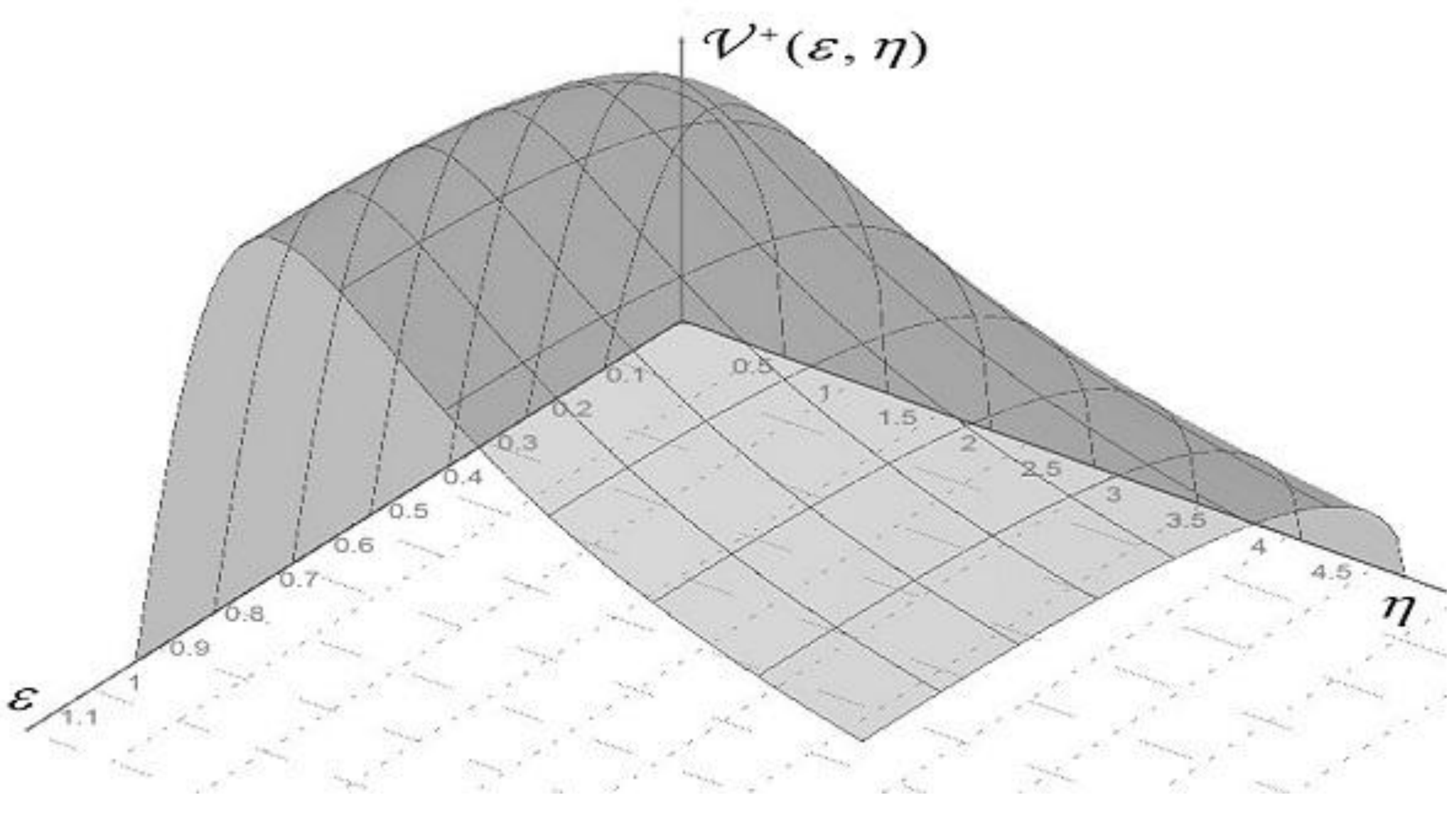

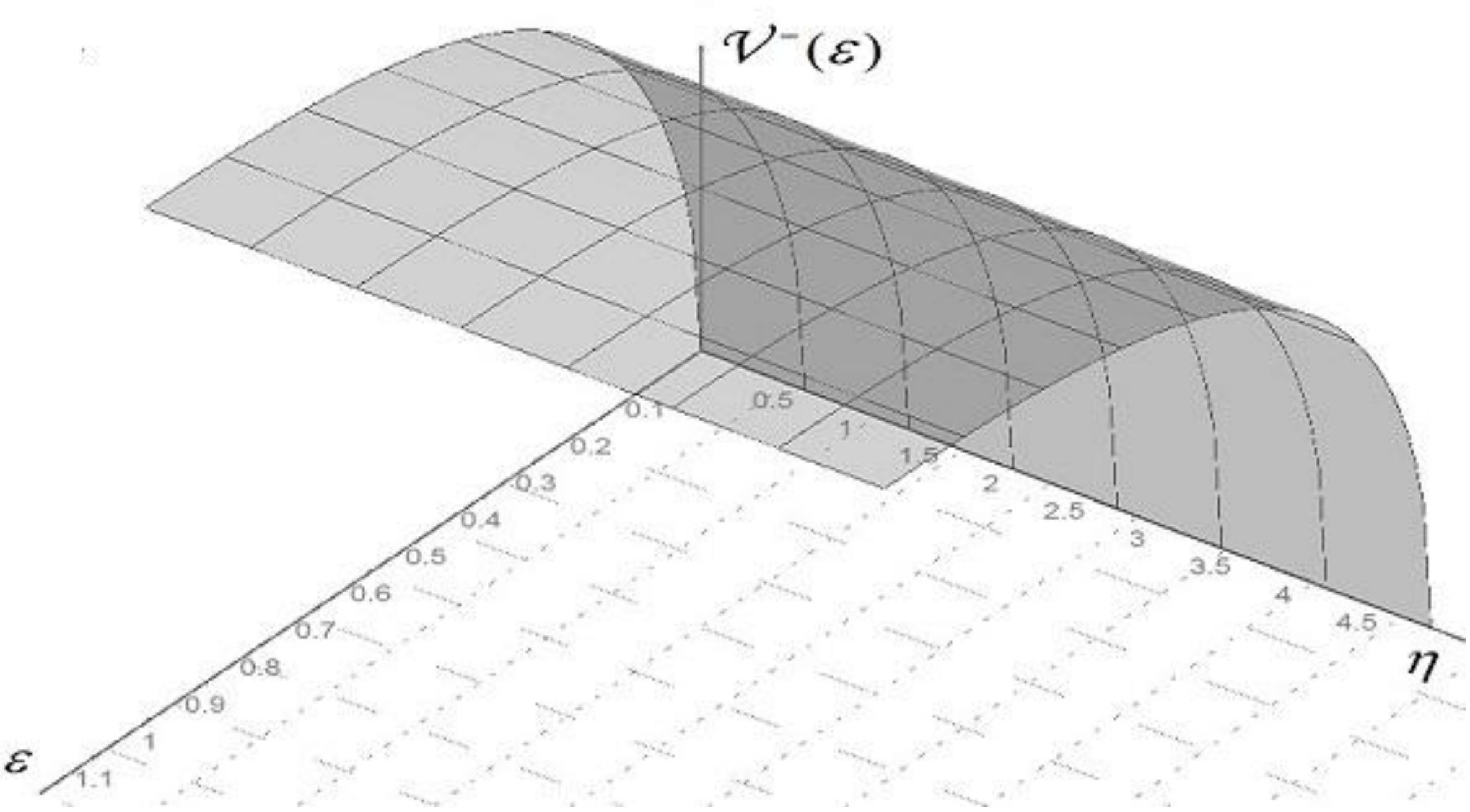

**Fig. 3**. Visibilities of interference patterns of (+) and (--) correlated states.

The contrast in (2.14) is, as mentioned above, maximal at $\varepsilon = \eta = 1$. Decreasing $\varepsilon$ from $\varepsilon = 1$ weakens the entanglement by making one of the superposed states in (2.1) more probable than the other, thus bringing each particle closer to a definite state [6].

Now turn to the *local* probabilities. The probability for photon A to hit detector $\mathcal{F}$A regardless of what happens to B is given by the sum of (2.10) and one half of (2.13):

$$\mathcal{P}_A(\mathcal{F}) = \mathcal{P}(\mathcal{F}, \mathcal{F}) + \mathcal{P}(\mathcal{F}, \mathcal{G}) = p^2 t^4 + q^2 r^4 + r^2 t^2 = \frac{1 + \varepsilon\eta}{(1+\varepsilon)(1+\eta)} \qquad (2.16)$$

Similarly, for probability of hitting detector $\mathcal{G}$ A we obtain

$$\mathcal{P}_A(\mathcal{G}) = \mathcal{P}(\mathcal{G}, \mathcal{G}) + \mathcal{P}(\mathcal{G}, \mathcal{F}) = p^2 r^4 + q^2 t^4 + r^2 t^2 = \frac{\varepsilon + \eta}{(1+\varepsilon)(1+\eta)} \qquad (2.17)$$

The same results hold for photon B as well, which allows one to drop the label A in (2.16, 17). The most important result is that both *local* probabilities (2.16, 17) are, in contrast with global interference (2.12, 13), phase-independent! So there is no local interference for *any* path-entangled photon pair. The generalized scheme reveals that even an arbitrarily weak entanglement completely "kills" local coherence. Mathematically, *local* coherence is zero in all domain of $\varepsilon$, $\eta$ and can exist only for totally disentangled photons. The BS-s eliminate the "which path" information, thus inviting local interference; but already arbitrarily weak entanglement kills the sensitivity of constituents to phases while sensitizing to them the whole system. The local coherence transfers in all its wholeness to the global scale.

Since basis $\left(|1\rangle, |2\rangle\right)$ is mathematically equivalent, e.g., to the electron spin states $\left(|\uparrow\rangle, |\downarrow\rangle\right)$, the obtained results should also hold for an entangled electron pair, even though the underlying physics is quite different [6]. Generally, coherence transfer from local to global scale must be a fundamental effect common at least for all entangled qubit pairs.

## Conclusion

The proposed changes in experimental setups of RTO open a possibility to study coherence transfer from local to global scale at arbitrary entanglement strength. The most important result can be formulated as the "mutual intolerance" between the local coherence of a bipartite members and their entanglement. In contrast with incompatible observables like position and momentum, which can still have definite expectation values in one state under indeterminacy relationship, there is no coexistence for local and global coherence.

It remains to perform the suggested thought experiment in order to verify (or refute) the predicted phenomenon.

## Acknowledgements

I am grateful to Art Hobson for stimulating discussions that had inspired this article.
I also want to thank Nick Herbert and Anwar Shiekh for their constructive comments.